\documentclass[ss]{imsart}
\usepackage{graphicx}
\usepackage{amsthm,amsmath}
\usepackage{amsfonts}
\RequirePackage[numbers]{natbib}
\RequirePackage[colorlinks,citecolor=blue,urlcolor=blue]{hyperref}
\usepackage{caption}
\usepackage{subcaption}
\usepackage{url}

\doi{10.1214/154957804100000000}
\pubyear{0000}
\volume{0}
\firstpage{1}
\lastpage{1}

\startlocaldefs
\newcommand{\bc}{\boldsymbol c}

\endlocaldefs

\begin{document}

\begin{frontmatter}

\title{Core-periphery structure in networks: a statistical exposition}
\runtitle{Core-periphery structure in networks}




\author{\fnms{Eric} \snm{Yanchenko}\ead[label=e1]{ekyanche@ncsu.edu}}
\and
\author{\fnms{Srijan} \snm{Sengupta}\thanksref{t3}\ead[label=e2]{ssengup2@ncsu.edu}}

\address{Department of Statistics\\
North Carolina State University\\
SAS Hall, 2311 Stinson Dr\\
Raleigh, NC 27607\\
\printead{e1,e2}}

\thankstext{t3}{Corresponding author}

\runauthor{Yanchenko and Sengupta}

\begin{abstract}
\noindent
Many real-world networks are theorized to have {\it core-periphery structure} consisting of a densely-connected core and a loosely-connected periphery. While this phenomenon has been extensively studied in a range of scientific disciplines, it has not received sufficient attention in the statistics community. In this expository article, our goal is to raise awareness about this topic and encourage statisticians to address the many open inference problems in this area. To this end, we first summarize the current research landscape by reviewing the metrics and models that have been used for quantitative studies on core-periphery structure.
Next, we formulate and explore various inferential problems in this context, such as estimation, hypothesis testing, and Bayesian inference, and discuss related computational techniques. 
We also outline the multidisciplinary scientific impact of core-periphery structure in a number of real-world networks.
Throughout the article, we provide our own interpretation of the literature from a statistical perspective, with the goal of prioritizing open problems where contribution from the statistics community will be most effective and important.
\end{abstract}





\end{frontmatter}

\section{Introduction}\label{sec:intro}

\begin{figure}
    \centering
    \includegraphics[width=\textwidth]{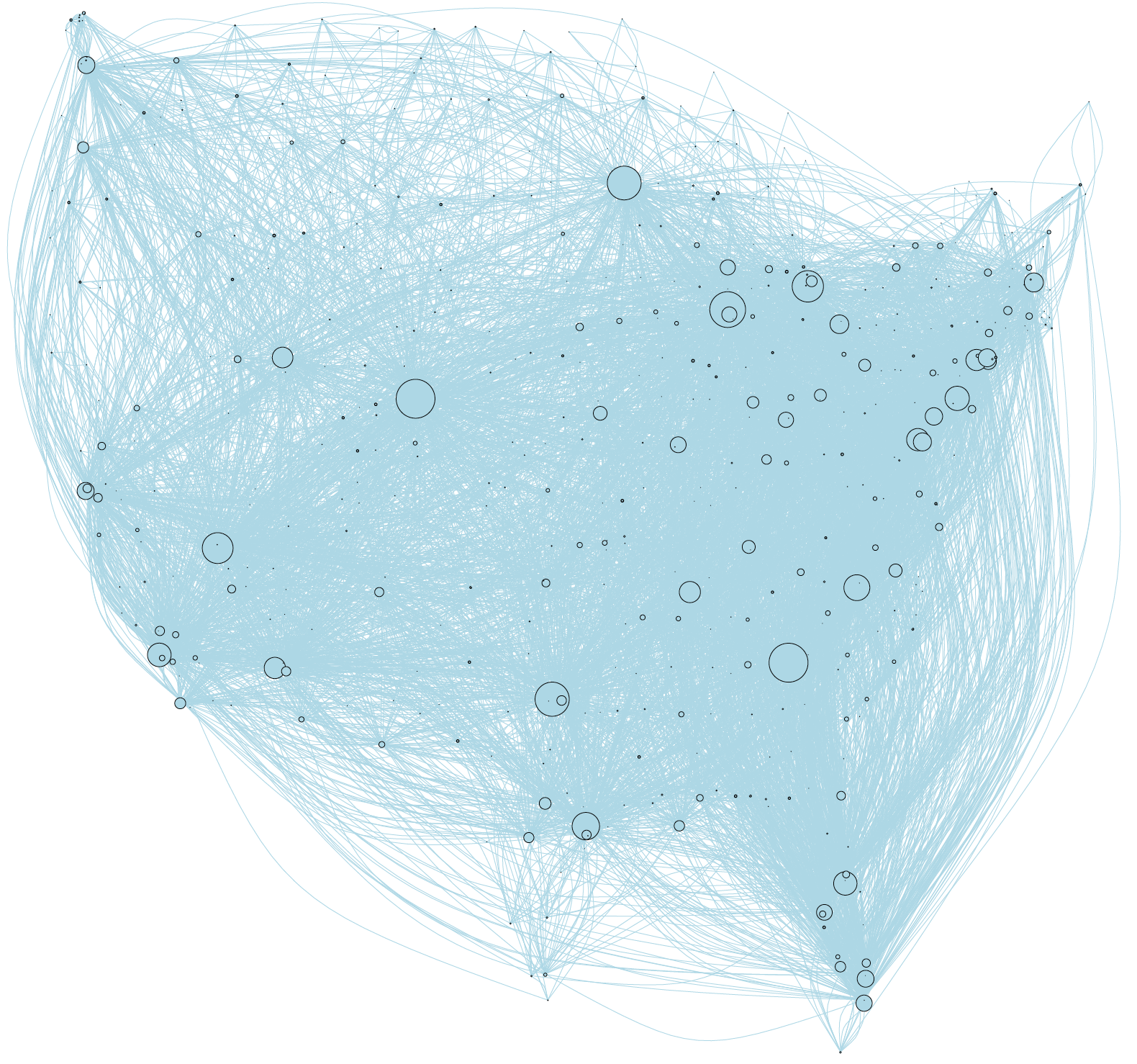}
    \caption{Airport network of the contiguous United States where each node is an airport and an edge represents a flight between two airports \cite{igraphdata}. The size of each node corresponds to its degree.}
    \label{fig:airports}
\end{figure}

 In its simplest form, a network\footnote{Throughout this work, we use the term ``network" to refer the mathematical object that is also known as a ``graph" or ``network-graph." The term ``network" can also mean the collection of interacting entities in the real world, but the meaning should be clear from context.} is a mathematical representation of a set of objects (e.g., social actors, entities) which interact (i.e., have some relationship) with each other. The interacting objects are called {\it nodes} and their interactions are called {\it edges}. 
In our highly-interconnected world, networks represent a powerful model to study complex systems.
As a result, many fields study networks, e.g., social interpersonal networks, where each node is a person and each edge represents some interaction like friendship or social contact \cite{scott1988social, mcpherson2001birds, kane2014s, dasgupta2022scalable, guo2020online}; infrastructural networks like airport networks, where each node is an airport and each edge represents a flight between them, as in Figure \ref{fig:airports} \cite{guimera2004modeling, li2004statistical}; citation networks where each node is a paper and each edge represents a citation \cite{lehmann2003citation, radicchi2008universality, bradley2020co, chandrasekharan2021finding}; and biological networks where each node is a cell or molecule and each edge represents an interaction in a biological process \cite{girvan2002community,alm2003biological, michailidis2012statistical}.

Statistics is a key component of today's data-driven science of networks.
Quoting from a high-profile joint editorial in the multidisciplinary journal {Network Science}: 
 ``\textit{Statistics is often defined as the study of data, involving anything from its collection, preparation, and management to its exploration, analysis, and presentation.
 In this view, 
our definition of network science delineates a subarea of statistics
concerned with data of a peculiar format}'' \cite{brandes2013network}.
Fittingly, the last two decades have seen excellent development of the statistical literature on networks, covering several aspects of statistical analysis, such as: data-generating models \cite{hoff2002latent,handcock2007model, noroozi2021hierarchy,senguptapabm};  consistency of estimation \cite{bickel2009nonparametric,zhao2012consistency, lei15}; spectral decomposition \cite{ng2002spectral, von2008consistency, sengupta2015spectral}, anomaly detection \cite{priebe2005scan,kodali2019value,jeske2018statistical}, and hypothesis testing \cite{bickel2016:aa, bhadra2019bootstrap, lovekar2021testing, yanchenko2021}.

Statistical formalism is especially important in the {\it meso-scale} or medium-scale analysis of networks, i.e., analyzing the properties of groups of nodes, instead of analyzing individual nodes (local-scale) or the entire network (system-scale). 
In the statistics literature, by far the most studied meso-scale property is {\it community structure} \cite{Newman:2004aa, Newman:2006aa, Fortunato:2010aa}. A community is a group of nodes, and a network is considered to have a community structure if nodes are heavily connected within communities but only loosely connected between communities, implying that nodes in the same community share certain fundamental characteristics.

The topic of this review is another meso-scale feature, {\it core-periphery} (CP) structure \cite{BORGATTI2000, Csermely2013}, which appears prominently in network science but has received little attention from the statistics community. 
Under this property, the set of nodes consists of two groups, a {\it core} and a {\it periphery}. 
Core nodes are densely connected to each other as well as to periphery nodes, whereas periphery nodes are only sparsely inter-connected. An alternative definition is that core nodes are a short distance from all other nodes. 
CP structure has been observed in a wide variety of real-world networks.
For example, in global trading networks, countries with large economies trade with both large and small economies, forming the core, while small economies are less likely to trade with each other, forming the periphery \citep{krugman1996self, magone2016core}. 
In airport networks, major airports (corresponding to large cities or airline hubs) have flights to other major airports as well as regional airports, but smaller airports have few flights between themselves \citep{lordan2017analyzing, lordan2019core}. 
Academic citation networks also exhibit a CP structure as high-profile papers receive citations from many types of papers whereas obscure papers are less likely to cite each other \citep{zelnio2012identifying, sedita2020invisible, wedell2022center}. In each of these examples, the groups of nodes (representing the core or the periphery) share some fundamental, underlying characteristic, which makes their assignment to the correct group an important task.

Studying the CP structure of a network can be important for understanding individual nodes as well. Nodes in the core are likely more ``influential" to the network in some sense. For example, in a power grid network, core power plants are more vital to the health of the grid than those in the periphery \citep[e.g.,][]{yang2021optimizing}. If a core power plant stops being operational due to a storm or a nefarious actor, the consequences on the power grid are much greater than if a peripheral plant went offline. Additionally, nodes on the boundary between the core and periphery may play a unique, mediatory role between the two groups \citep{cattani2008}. 
Understanding the CP structure of a network helps the researcher determine which nodes are the most significant and worthy of further investigation.

 \begin{figure}
    \centering
        \includegraphics[width=0.45\textwidth, trim={2.5cm 2.5cm 2cm 2cm},clip]{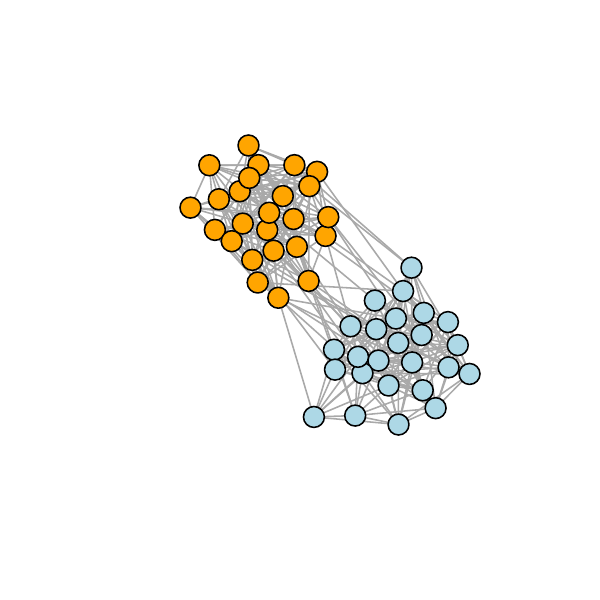} 
        \includegraphics[width=0.45\textwidth, , trim={2cm 2cm 2cm 2cm},clip]{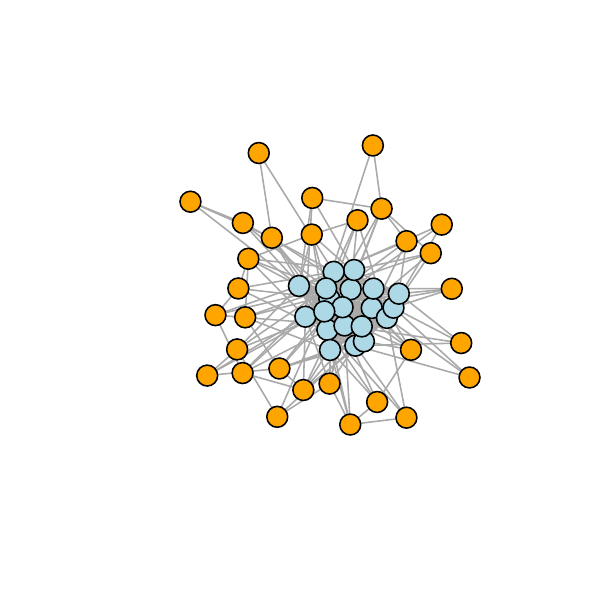} 
     \caption{Comparison of networks generated with a community structure (left) versus a core-periphery structure (right).}
    \label{fig:comm_cp}
\end{figure}

 There are some key distinctions between a CP structure and community structure.
 Community structure implies
 low edge density between communities and high edge density within communities.
 In contrast, a CP structure implies two groups, typically one small group (core) and one large group (periphery), with high edge density within the first group, medium edge density between the two groups, and low edge density within the second group. 
 Under community structure, all nodes prefer within-group connections over between-group connections.
 Under a CP structure,  core nodes prefer within-group connections over between-group connections, but periphery nodes  prefer between-group connections over within-group connections.
  Another fundamental difference between these two structures is that a network with communities can be broken down into separate, minimally interacting, self-contained sets of nodes. This is in contrast to a CP structure where core nodes influence and interact with the entire network. In Figure \ref{fig:comm_cp}, we see how community structure is composed of two densely connected sets of nodes which have minimal edges between them. A CP structure, alternatively, has a dense core (blue nodes) that is highly connected to the periphery (orange nodes) but has few intra-periphery edges. It is also possible for a network to have multiple communities and/or CP structures.

The goal of this review is to survey the research landscape of CP structures from a statistical perspective.
We are not only interested in methods to identify this network feature, but also in understanding how these structures are generated, their significance and what they mean in the context of the specific application. Our work builds off the reviews in \cite{Csermely2013} and \cite{tang2019} but has a greater emphasis on important statistical concepts related to CP structures, such as hypothesis testing and Bayesian inference. Another contribution of this review is that, in each section, we include a careful discussion of the existing methods, comparing and contrasting them and presenting the strengths and weakness of the literature in this area. We begin this work by introducing standard notation and random graph models. In Section \ref{sec:met}, we describe a number of CP metrics from the network science literature. Statistical inference tasks for a CP structure are covered in Section \ref{sec:inf}, where we consider generative models, consistency, hypothesis testing, and Bayesian inference. Sections \ref{sec:comp} and \ref{sec:data} explore computational techniques and real-world examples, respectively. We conclude in Section \ref{sec:open} by proposing important avenues for future research.

\subsection{Network basics}
We begin with a brief overview of notations and statistical models used to describe networks. 
Consider a network of $n$ nodes and let $A$ be its $n\times n$ \textit{adjacency matrix}. Then $A_{ij}=1$ if nodes $i$ and $j$ are connected by an edge and $A_{ij}=0$ otherwise. If network interactions are mutual in nature, e.g., being friends on Facebook, then we say that the network is {\it undirected} and $A_{ij}=A_{ji}$ for all $i\neq j$. 
If network interactions are not mutual in nature, e.g., a paper citing another paper, then $A_{ij}=1$ if there is an edge {\it from} node $i$ {\it to} node $j$ and we say that the network is {\it directed}. 
If a node has an edge with itself, e.g., sending an email to oneself, then we call this a {\it self-loop} and $A_{ii}=1$. 
Lastly, edges need not be binary and can instead have a {\it weight}. A {\it weighted} network allows edges to take values in $\mathbb R$, e.g., an airport network where the nodes are airports and the weight of an edge is the number of daily flights between airports.
Unless otherwise noted, for the remainder of this work, we only consider {\it simple} networks which are unweighted, undirected networks and contain no self-loops, and we define $m=\sum_{i<j}A_{ij}$ to be the total number of edges. Note that any network can be transformed to a simple network by symmetrizing the edges and removing weights and self-loops.

Now, let us consider data generating models for simple networks.
Let $A\sim P$ mean that $A_{ij}\sim\mathsf{Bernoulli}(P_{ij})$ independently for $1 \le i < j \le n$.
Under this general framework, a number of statistical models have been proposed with different specifications of $P$, the data-generating matrix.
The simplest such model is the {\it Erd\"{o}s-R\'{e}nyi} (ER) model \cite{erdos1959} where $P_{ij}=p$ for all $i\neq j$, i.e., each node has the same probability, $p$, of connecting with each remaining node. While this is a very simple model and unlikely to be observed in reality, it can be useful as a benchmark or null model. A generalization of the ER model to allow for heterogeneous edge probabilities is the Chung-Lu (CL) model \cite{chung2002average}, where the model parameters consist of a {\it weight vector} $\boldsymbol\theta=(\theta_1,\dots,\theta_n)$, and we have $P_{ij}\propto\theta_i\theta_j$. Nodes with larger $\theta_i$ are more likely to have edges in general, and, in particular, more likely to have edges with other nodes with large weights. A CL model may make sense for networks with {\it degree heterogeneity} where the {\it degree} of a node is the number of edges for that node, defined as $d_i=\sum_{j=1}^n A_{ij}$ for node $i$. The degree of a node is useful and simple measure of node importance.

To generate networks with community structure, a standard approach is to use a stochastic block model (SBM) \cite{holland83}. Let $\boldsymbol c=(c_1,\dots,c_n)$ be the community assignment vector where $c_i=b$ means that node $i$ is in community $b\in\{1,\dots,B\}$, and $B$ is the number of communities. Let $\Omega$ be the $B\times B$ connectivity matrix where $\Omega_{uv}$ is the probability of an edge between a node in community $u$ and community $v$, and we have $P_{ij}=\Omega_{c_ic_j}$. The name ``block" comes from the fact that the adjacency matrix has an approximate block structure since nodes belonging to the same community are stochastically equivalent. A generalization of the SBM is the Degree-Corrected Stochastic Block Model (DCSBM) \cite{karrer2011stochastic} which combines the ideas of the CL model and the SBM. Using the same notation as in previous models, we have $P_{ij}=\theta_i\Omega_{c_ic_j}\theta_j$. This model keeps the same block structure as an SBM but allows for ``hubs'', i.e.,
nodes with high degree. 
The popularity-adjusted blockmodel \cite{senguptapabm, noroozi2020statistical} further extends the DCSBM to allow nodes to be central in one community but peripheral in other communities.
Throughout this paper we will consider how each of these data-generating models apply to CP structures.

\section{Metrics}\label{sec:met}
In this section, we review metrics that are already popular in the scientific literature for quantifying the CP structure of a network.
We believe that examining these approaches statistically can make significant contributions to the CP literature, and it is therefore helpful for statisticians to become familiar with them.

One of the primary challenges of studying CP structures is that a universal, formal definition does not exist. This is akin to the problem of defining communities or clusters in multivariate data analysis. Because of this, there are many descriptors to identify and quantify a CP structure, but the general form is that of a highly-connected {\it core} with a loosely-connected {\it periphery} \cite{BORGATTI2000, Csermely2013}. Each method implicitly defines a different CP structure, and most of these definitions can be categorized into two frameworks: the block-model framework and the transport framework. Another important point to consider when defining a CP structure is whether the core-to-periphery edges are dense or not (most authors take it to be dense).  Additionally, the core size should be much smaller than the rest of the network. If, for example, 95\% of the nodes were in the core, this should not constitute a CP structure. 

There are several other concepts that are closely related to CP structures. Rich clubs \cite{zhou2004rich} are characterized by interconnected high-degree nodes and bow-tie structures for directed networks \cite{ma2003connectivity, supper2009bowtiebuilder} have core nodes with many incoming and outgoing edges. For the sake of time and space we limit this discussion to CP structures. Lastly, we define a ``core-periphery assignment'' to be an integer-valued vector $\bc$ of length $n$, such that $c_i=1$ if node $i$ is in the core and $c_i=0$ otherwise. We define the size of the core as $K=\sum_{i=1}^n c_i$.

\subsection{Block model metrics}
First, we consider metrics which assume a block CP structure, i.e., the adjacency and/or data-generating matrix have a block structure. The defining characteristic of this approach is that the core is dense and the periphery is sparse. Borgatti and Everett \cite{BORGATTI2000} took the first, and by far most cited, foray into studying CP structures. Their method measures how well the observed network agrees with an ``ideal" CP structure. In particular, let $\Delta(\bc):=\Delta$ be the ideal CP structure where $\Delta_{ij}=1$ if node $i$ or node $j$ is in the core, and $\Delta_{ij}=0$ otherwise, i.e., $\Delta_{ij}=c_i+c_j-c_ic_j$. Then the Borgatti and Everett metric is
\begin{equation}\label{eq:rho}
    \rho = \mathsf{Cor}(A,\Delta)
\end{equation}
where $\mathsf{Cor}(\cdot,\cdot)$ is the Pearson correlation.

If the CP assignments are known {\it a priori}, then one simply computes $\rho$ for this assignment. For example, in a world trade network where countries are nodes and edges represent inter-country trade, we might consider the nations with large economies as the core  and compute $\rho$ with these labels \citep{kostoska2020core}. More often, however, we do not have any information on the nodes so we want to determine whether {\it any} assignment has a CP structure. In this more realistic and challenging case, $\rho$ is maximized over the space of all possible assignments using a combinatorial optimization routine such as genetic algorithms \cite{anderson1994, sastry2005}. See Section \ref{sec:comp} for further discussion on algorithms. 

To interpret $\rho$ from a statistical perspective, we write it out in more detail to see that
\begin{equation}
    \rho 
    = \frac{M_{\bc}(K)-a_K}{b_K}, \text{ where }
\end{equation}
\begin{equation}\notag
    M_{\bc}(K) = \sum_{i<j} A_{ij} \Delta_{ij},\
    a_K = {n\choose 2} \text{mean}(A)\text{mean}(\Delta),\ 
    b_K =\{\text{var}(A)\text{var}(\Delta)\}^{1/2}
\end{equation}
where $M_{\bc}(K)$ is the number of core-core and core-periphery edges and $\text{mean}(\cdot)$ and $\text{var}(\cdot)$ are the sample mean and variance operators, respectively. Conditional on $A$, $a_K$ and $b_K$ depend only on the number of nodes in the core, $K$, and increase as $K$ increases. Thus, maximizing $\rho$ over the set of CP assignments is equivalent to maximizing the number of core-core and core-periphery edges with a penalty for the size of the core. This definition agrees with our understanding of a CP structure since it yields large values for networks with a dense core and sparse periphery. It also means that core-core and core-periphery edges are treated equivalently. This descriptor is statistically tractable because it can be easily computed and has a natural interpretation with respect to the adjacency matrix. Other authors have generalized this work \cite[e.g.,][]{Rombach2017, kojaku2017} but we omit the discussion of these extensions here.


In a similar vein, Brusco \cite{BRUSCO2011} finds the CP labels which maximize
\begin{equation}\label{eq:brusco}
    Z(A, \bc)
    =\sum_{i<j,c_i=c_j=1} \mathbb I(A_{ij}=1) +
    \sum_{i<j,c_i=c_j=0} \mathbb I(A_{ij}=0)
    \footnote{In the original paper, they minimize $Z'(\bc)
    =\sum_{i<j,\bc_i=\bc_j=1} \mathbb I(A_{ij}=0) +
    \sum_{i<j,\bc_i=\bc_j=0} \mathbb I(A_{ij}=1)$ but we note that these are equivalent formulations} 
\end{equation}
where $\mathbb I(Q)=1$ if $Q$ is true and 0 if false. In other words, this method maximizes the number of core-core edges while minimizing the number of periphery-periphery edges. A perfect CP structure occurs when every core node has an edge with every other core node and there are no edges between periphery nodes. This method is similar to that of Borgatti and Everett in that it maximizes core-core edges with a ``penalty" term, but here the penalty is the number of periphery-periphery edges, as opposed to the core size. Additionally, (\ref{eq:brusco}) does not account for core-periphery edges.

For an approach with more of a statistical flavor, we consider Zhang et al. \cite{Newmann2015}. The authors consider an SBM with two blocks where $p_{00},p_{11}$ and $p_{01}$ are the intra-community probability of an edge for communities 0 and 1 and the inter-community probability of an edge, respectively. Nodes are randomly assigned to community 0 with probability $\gamma_0$ and community 1 with probability $\gamma_1=1-\gamma_0$. Then the likelihood is
\begin{equation}\label{eq:like}
    P(A|p,\gamma)
    =\sum_{\bc}\prod_{i<j} p_{c_ic_j}^{A_{ij}}(1-p_{c_ic_j})^{1-A_{ij}}\prod_i \gamma_{c_i}
\end{equation}
where the sum is taken over all possible CP assignments, $\bc$.
This expression is then maximized over $p_{ij}$ and $\gamma_k$ using an Expectation-Maximization (EM) algorithm \cite{dempster1977maximum}. This optimization problem yields $q_r^i$, the probability that node $i$ is assigned to group $r\in\{0,1\}$, as well as estimated probabilities $\hat p_{00},\hat p_{01},\hat p_{11}$. Mathematically,
$$
    q_r^i
    =\sum_{\bc} q(\bc)\mathbb I(c_i=r)
$$
where
$$
    q(\bc)
    =\frac{\prod_{i<j} p_{c_ic_j}^{A_{ij}}(1-p_{c_ic_j})^{1-A_{ij}}\prod_i \gamma_{c_i}}{\sum_{\bc'}\prod_{i<j} p_{c_i'c_j'}^{A_{ij}}(1-p_{c_i'c_j'})^{1-A_{ij}}\prod_i \gamma_{c_i'}}
$$
To assign binary CP labels $\bc$, each node is assigned to the group for which it has a higher probability of membership, i.e., $c_i=\mathbb I(q_1^i > 0.5)$. Thus, this method yields a continuous model for CP labeling.

Now, if $\hat p_{11} > \hat p_{01} > \hat p_{00}$ then these labels correspond to a CP structure as this result directly relates to the CP definition. If the likelihood is maximized with $\hat p_{11}>\hat p_{00}>\hat p_{01}$, then this is strong evidence in favor of community structure and against a CP structure. Therefore, this approach yields a sense of the statistical significance of the CP structure. This model, however, requires that the inter- and intra-community probabilities be homogenous, something likely violated in practice. Additionally, the maximization of (\ref{eq:like}) could return arbitrarily large cores, but this rarely occurs in practice.

Next, we consider a method which builds on the seminal work of Caron and Fox \cite{caron2017sparse}. In \cite{caron2017sparse}, the authors consider networks from a point-process framework in order to achieve sparsity as well as exchangeability. Let $Z$ be a point-process on $\mathbb R^2_+$ and
\begin{equation}
    Z
    =\sum_{i,j} z_{ij}\delta_{t_i,t_j}
\end{equation}
where $z_{ij}=1$ if there is an edge between nodes $i$ and $j$, $\delta_{x,y}$ is the Dirac delta function which takes the value $\infty$ at $(x,y)$ and 0 elsewhere with $\int \delta_{x,y}\ dx\ dy=1$ and $t_i,t_j\in\mathbb R_+$ represent a kind of time index. Each node has a {\it sociability parameter } $\theta_i>0$ (similar to the weights in a CL model) such that the probability of an edge between nodes $i\neq j$ is
\begin{equation}\label{eq:pp}
    P(z_{ij}=1|t_i,t_j,\theta_i,\theta_j)
    =1-e^{-2\theta_i\theta_j},
\end{equation}
independent of the time indices. The authors further develop the theory and mechanics using Completely Random Measures \cite{kingman1967completely}. 

Naik et al. \cite{naik2021} take this novel approach of modeling networks and apply it to CP structures. The authors define a CP structure as a sparse network with dense core sub-network. A network is said to be {\it dense} if the number of edges $m=O(n^2)$ and {\it sparse} if $m=o(n^2)$. The probability of an edge between two nodes is formulated to be a slight generalization of (\ref{eq:pp}). Specifically, 
\begin{equation}
    P(A_{ij}=1|\boldsymbol\theta)
    =1-e^{-2(\theta_{i1}\theta_{j1} +\theta_{i2}\theta_{j2})}
\end{equation}
where $\theta_{i1}\geq 0$ is a core parameter and $\theta_{i2}$ is the core-periphery parameter. The model parameters $(\theta_{i1}, \theta_{i2},t_i)_{i=1,\dots,n}$ are the points from a Poisson process on $\mathbb R^{3}_+$ with mean measure $\nu(d\theta_1,d\theta_2)dt$ where $\nu$ is a $\sigma$-finite measure on $\mathbb R^2_+$, concentrated on $\mathbb R^2_+\setminus\{(0,0)\}$, where $\int_{R^2_+} \min(1,\theta_1+\theta_2)\nu(d\theta_1,d\theta_2)$ is finite. The network is sparse if
\begin{equation}
    \int_{\mathbb R^2_+}\nu(d\theta_1,d\theta_2) = \infty
\end{equation}
and dense otherwise. Using compound Completely Random Measures \cite{griffin2017compound}, the authors achieve $\theta_{i1}\geq 0$ and $\theta_{i2}>0$. Then a CP structure is enforced with the following assumptions on $\nu$:
\begin{align}
    \int_{(0,\infty)\times \{0\}} \nu(d\theta_1,d\theta_2) &=0\\
    \int_{\{0\}\times (0,\infty)} \nu(d\theta_1,d\theta_2) &>0\\
    0<\int_{(0,\infty)^2} \nu(d\theta_1,d\theta_2) &<\infty.
\end{align}
The nodes with $\theta_{i1}>0$ are labeled as core nodes and $\theta_{i1}=0$ as periphery nodes. We refer the reader to the paper for full details.

By characterizing a CP structure as a dense core within a sparse graph, the authors implicitly enforce the core to be small, since a large core would make the entire network dense. The concepts of dense/core graphs are also intuitive and connected to conceptual definition of a CP structure. This approach, however, does not allow for the core-periphery edges to be dense and there is no explicit estimator, making comparison to other methods difficult. 

\subsection{Transport metrics}

We now consider transport-based methods where the formulation depends on distance or centrality measures.
The central idea is that core nodes are a short distance from all other nodes. 
We start with the work of Holme \cite{Holme2005}. The author states that for a network to have a CP structure it must have a high {\it closeness centrality} in addition to a well-defined cluster. The closeness centrality \cite{sabidussi1966centrality} of a node is a well-known measure of centrality, defined as the inverse of the average distance between one node and all other nodes. One can also find the closeness centrality of a subset of nodes. Let $V$ be the set of all nodes of a network and $U\subseteq V$ be a subset of nodes with $|U|=n_U$. Then the closeness centrality of $U$, $C_C(U)$, is defined as
\begin{equation}
    C_C(U) = \left(\frac1{n_U(n-1)}\sum_{i\in U}\sum_{j\in V\setminus\{i\}} d_{ij}\right)^{-1}
\end{equation}
where $d_{ij}$ is the distance between node $i$ and $j$ (number of edges it takes to go from node $i$ to node $j$). If the average distance between nodes in $U$ and all other nodes is small, then the terms in the sum will be small. Taking the inverse will then yield a large value of $C_C(U)$ and thus the nodes in $U$ are deemed to be more important.  On average, a short distance between nodes is a desirable characteristic of a core. Now, consider a sequence of potential cores, $V_1,V_2,\dots$, where $V_k$ is the {\it k-core}, defined as the maximal subset of nodes such that each node is connected to at least $k$ other nodes in the subset \cite{wasserman1994social, newman2018networks}. The core, $V_{core}$, is the $k$-core that maximizes the closeness centrality, i.e., $V_{core}=\arg\max_{V_k}\{C_C(V_k)\}$. Then the proposed CP metric, $c_{cp}$, is defined as
\begin{equation}\label{eq:holme}
     c_{cp}
    =\frac{C_C(V_{core})}{C_C(V)} - \frac1{|\mathcal G|}\sum_{G'\in\mathcal G} \frac{C_C(V'_{core})}{C_C(V')}   
\end{equation}
where $\mathcal G$ is a set of networks with the same degree sequence as the observed network. This ``null model" is also called the {\it configuration model} \cite{newman2018networks}.

There are several note-worthy aspects of this approach. First, the definition of the core is intuitive and computationally efficient. Next, the first term in (\ref{eq:holme}) is ``normalized" by the closeness centrality of the entire network. This means that if the entire network is well-connected, then the closeness of the core will not be significantly greater than the closeness of the network, making this term smaller. Thus, the core must be be significantly more well-connected than the rest of the network. This also ensures that the core is small since otherwise this term would be close to one. Lastly, Holme's metric is compared against a ``baseline" value. By averaging over an ensemble of networks with the same degree sequence as the original network, $c_{cp}$ gives a proxy for the significance of this structure in the network. A positive value of $c_{cp}$, therefore, signifies that there is a CP structure greater than what would be expected by a random network with the same degree sequence. Conversely, a negative value indicates the CP structure may simply be the result of noise in the data. There are, however, no statistical guarantees for this procedure and it is unclear whether the degree-preserving random network is a reasonable null model. We return to the topic of null models in Section \ref{sec:inf}. A drawback to this approach, as well as all ensuing transport methods, is that it can be difficult to develop statistical theory. For example, finding the moments of $c_{cp}$ is highly non-trivial as it depends on $k$-cores and closeness centrality, both of which are non-linear functions of the adjacency matrix. Indeed, any metric that depends on graph distances is a complicated function of the adjacency matrix.

Another transport-based approach is from da Silva et al. \cite{da2008}. The authors first introduce the idea of the {\it capacity} of a network to measure its overall connectedness. Let $d_{ij}$ be the distance between nodes $i$ and $j$.  Then the capacity of the network, $\mathcal C$, is
\begin{equation}
    \mathcal C
    =\sum_{i<j} \frac1{d_{ij}}, 
\end{equation}
such that many short paths paths lead to a larger capacity. da Silva et al. argue that a core is well-connected to the rest of the graph in the sense that the removal of a core node substantially reduces the network's capacity, $\mathcal C$. Based on empirical findings, the authors define the core-coefficient, $cc$, as $cc=N/n$ where $N$ satisfies
\begin{equation}\label{eq:da}
    \sum_{i=0}^N \mathcal C_i = 0.9\sum_{j=0}^n \mathcal C_j,
\end{equation}
and where $\mathcal C_i$ is the capacity of the network after removing $i$ nodes in decreasing order of closeness centrality. If the capacity substantially decreases with the removal of certain nodes, then these nodes proved to connect many other nodes in the network. In an airport network, for example, if CLT airport (Charlotte), the operational hub for American Airlines, became non-operational due to a hurricane, then many more connecting flights would be disrupted as compared to if RDU (Raleigh-Durham) airport became non-operational. Thus, CLT is a core node and RDU is not. The authors also suggest a threshold of $cc>0.5$ to signify that that the network exhibits a CP structure that is more pronounced than what might otherwise be expected by chance alone. This, along with the 90\% threshold in (\ref{eq:da}), however, were arbitrarily selected based on a small empirical study and not justified theoretically. This approach is similar to that of Holme \cite{Holme2005} in that it does not require any complicated optimization to detect the core, but rather chooses the nodes with largest closeness centrality. Additionally, da Silva et al. implicitly control for the size of the core. For example, if a network had a large, highly-connected core, then the entire network would necessarily be highly-connected as well. But then removing a single node would not reduce the capacity by a significant amount and thus the node would not be deemed a part of the core.

\subsection{Other metrics}

Lastly, we look at two methods which do not neatly fall into either the block model or transport paradigm. In Rossa et al. \cite{rossa2013}, the authors approach the CP identification problem from a random walk perspective. Consider a connected network and a random walk from node $i$ to node $j$ with probability $m_{ij}=A_{ij}/d_i$, i.e., jumps with equal probability from node $i$ to one of its neighbors. Let $\pi_i>0$ be the asymptotic probability of being at node $i$. Here, $\pi_i=d_i/2m$. Now, let $S\subseteq \{1,\dots,n\}$ denote the nodes in a sub-network. Then the {\it persistence probability} $\alpha_S$ is probability that the random walk, currently in $S$, stays in $S$ after its next jump. We find that
\begin{equation}
    \alpha_S
    =\frac{\sum_{i,j\in S} \pi_i m_{ij}}{\sum_{i\in S} \pi_i}
    =\frac{\sum_{i,j\in S} A_{ij}}{\sum_{i\in S}d_i}.
\end{equation}

The authors argue that under an ideal CP structure, we would have $\alpha_S=0$ when $S$ is the periphery group, since there would be no periphery-periphery edges. Thus, they seek to find the {\it $\alpha$-periphery} which is the largest sub-network $S$ such that $\alpha_S\leq \alpha$ for some $0<\alpha<1$. In other words, if the random walk is in an $\alpha$-periphery, then it will jump out of the sub-network at the ensuing step with probability $1-\alpha$. To do this, they consider an increasing sequence of sets $S_1,\dots,S_n$ where $S_1$ is the weakest connected node and each subsequent set adds the node which leads to the smallest increase in persistence probability, ending with $S_n$ being the entire network. This leads to a {\it core-periphery profile}, $0=\alpha_1\leq\alpha_2\leq\cdots\leq\alpha_n=1$ and the core is taken to be the largest set $S_k$ such that $\alpha_{S_k}\leq\alpha$.

This approach can also be used to obtain a quantification of the CP structure in the network called the {\it core-periphery centralization}, $C$. First, plot the core-periphery profile against the proportion of nodes in the core, yielding a curve. Then take the integral of this curve and compare it to an ``ideal" CP network (star graph) which has a value of 0 at $\alpha_{S_k}$ for $k=1,\dots,{n-1}$ and 1 for $\alpha_{S_n}$. Mathematically,
\begin{equation}\label{eq:rossa}
    C
    =1-\frac{2}{n-1}\sum_{k=1}^{n-1}\alpha_{S_k}.
\end{equation}
The larger the value of $C$, the closer the network is to the star graph and thus the more pronounced the CP structure is in the network.

Rossa et al. define the core as the nodes where the random walk ``spends the most time" or is most unlikely to leave, which is fundamentally different from other core descriptions. Another key difference with previous methods is that it finds the periphery and then takes the complement of that set to be the core, as opposed to finding the core and then setting the remaining nodes to the periphery. Similar to Holme and da Silva et al., however, it is difficult to compute standard statistical measures of (\ref{eq:rossa}) like its mean and variance.

Finally, we consider Cucuringu et al. \cite{cucuringu2016detection}. Here, each node is assigned a measure of ``coreness" based on the ideas of the {\it betweenness centrality}. Specifically, the {\it Path-Core} of node $i$ is defined as
\begin{equation}\label{eq:cuc}
    \sc{Path-Core}(i)
    =\sum_{j<k, \neq i, A_{jk}=1} \frac{\sigma_{jk}(i)|_{A_{jk}=0}}{\sigma_{jk}|_{A_{jk}=0}}
\end{equation}
where $\sigma_{jk}|_{A_{jk}=0}$ is the number of shortest paths from node $j$ to $k$ when $A_{jk}=0$, i.e., the edge from $i$ to $j$ is removed; and $\sigma_{jk}(i)|_{A_{jk}=0}$ is the number of shortest paths from node $j$ to $k$ which go through node $i$, when $A_{jk}=0$. Note that the sum is only over node pairs with edges, but the paths are computed between these nodes with their edge removed, thus looking at how close two connected nodes would be if the edge connecting them was removed. Then (\ref{eq:cuc}) is interpreted as the probability that the node is in the core. The intuition is that if many of the shortest paths in a network go through a particular node, then this node is likely part of the core. 

While (\ref{eq:cuc}) yields a measure of coreness of each node, in order to discriminate nodes as core or periphery, Cucuringu et al. suggest to find the labels which maximize the core-core and core-periphery edge density while minimizing the periphery-periphery edge density with a penalty for imbalanced core size. Mathematically, this maximizes
\begin{equation}\label{eq:cuc2}
    \xi(A,\bc)
    ={K\choose 2}^{-1}M_{cc}+\frac1{K(n-K)}M_{cp}-{n-K\choose 2}^{-1}M_{pp}
    -\gamma\left|\frac{K}{n}-\beta\right|
\end{equation}
where $M_{cc},M_{cp},M_{pp}$ are the number of core-core, core-periphery and periphery-periphery edges, respectively.
Here $\beta$ is the core size that is being shrunk towards and $\gamma$ is a tuning parameter. Instead of maximizing over all possible CP labels, the authors consider a much smaller space of solutions. Starting with the node with largest Path-Core score as the only core node, the algorithm adds nodes to the core in decreasing order of Path-Core score, yielding $n-2$ values of the objective function. Then the labels which correspond to the largest value of the objective function are kept as the core labels. 

Curcuringu et al. utilize both block model and transport ideas in their approach. A centrality measure is used to compute the Path-Core score while the objective function assumes a block model CP structure. The search space of the optimization problem is also greatly reduced from exponential in $n$ to linear. Another attractive aspect is that the maximum of $\xi(A,\bc)$ over $\bc$ estimates an intuitive statistical quantity, namely $\bar p_{cc}+\bar p_{cp}-\bar p_{pp}$, where $\bar p_{cc},\bar p_{cp},\bar p_{pp}$ are the average core-core, core-periphery and periphery-periphery edge probabilities, respectively. Lastly, the core size is explicitly controlled for by a penalization term.

\section{Statistical inference}\label{sec:inf}

Having studied some of the existing CP quantification methods, we now turn our attention to statistical inference for CP structures. This section focuses on four main aspects of a statistical inference framework: generative models, estimation, hypothesis testing, and Bayesian approaches. 

\subsection{Generative models}
The most common model to generate a CP structure is the block model. Consider a two block SBM where $p_{11},p_{00},$ and $p_{01}$ are the block-block edge probabilities,
i.e.,
\begin{equation}
    P(A_{ij}=1)
    =p_{c_i,c_j}
\end{equation}
where $c_i=1$ if node $i$ is in block 1 and 0 otherwise. If $p_{11}> p_{01}> p_{00}$, then the resulting network has a CP structure with community 1 being the core and community 0 being the periphery. This model is intuitive and simple, and thus is used in many works \cite[e.g.][]{Newmann2015}. 
However, this model might be considered simplistic due to two reasons.
First, the sharp boundary between the core and periphery is unrealistic. 
Second, the model requires perfect homogeneity within the core and periphery groups, which is unlikely to occur in real-world networks. 

Fox and Caron \cite{caron2017sparse} propose another generative model. Each node is given a {\it sociability parameter} $\theta_i>0$ and the probability of an edge between nodes $i\neq j$ is
\begin{equation}
    P(A_{ij}=1)
    =1-e^{-2\theta_i\theta_j}.
\end{equation}
Clearly, the larger $\theta_i$ and $\theta_j$, the greater the probability of an edge. 
To generate a network with a CP structure, we endow nodes in the core with a larger value of $\theta_i$ than those in the periphery. This naturally results in a dense core and a sparse periphery with the core-periphery edge density being in-between. Thus, this model attains a CP structure while also allowing for heterogeneous edge probabilities and a smoother transition between the core and periphery as compared to the block model. Jia and Benson \cite{Jia2019} describe a similar model where the edge probability is
\begin{equation}
    P(A_{ij}=1)
    =\frac{e^{\theta_i+\theta_j}}{1 + e^{\theta_i+\theta_j}}.
\end{equation}
Notice that, as a special case, if each node in the core has parameter $\theta_1$ and each periphery node has parameter $\theta_2$, then this model is equivalent to the SBM. The authors propose a further extension of this model which can account for spatial information of each node. Let $K_{ij}$ be an arbitrary (spatial) kernel function, e.g., $K_{ij}=||{\bf x}_i-{\bf x}_j||_2$, where ${\bf x}$ are spatial positions and $||\cdot||_2$ is the Euclidean distance. Then
\begin{equation}
    P(A_{ij}=1)
    =\frac{e^{\theta_i+\theta_j}}{K_{ij} + e^{\theta_i+\theta_j}}.
\end{equation}
This allows for node covariate information, if available, to be used to generate (or model) networks.  


The main difference between these approaches is that the block model enforces a hard cutoff between the core and periphery nodes whereas the other models allow a more flexible transition. 
The choice of model therefore depends on the user's definition of the boundary between the core and the periphery, and on the context of the network being studied. 
Under both modeling frameworks, higher degree nodes are core nodes and lower degree nodes are periphery nodes.
This raises a question on the interplay between node degrees and CP structure: at what point does the network simply have degree heterogeneity and at what point does it become a CP structure? This question is further investigated in Section \ref{sec:hyp}. 
\subsection{Estimation/Consistency}

An important question when estimating underlying or hidden labels in a block model is: did we recover the {\it correct} labels? That is, do the labels returned by the algorithm correspond to the ground-truth? 
This question relates to the statistical {\it consistency} of a method. 
Since this topic is well-studied for block models, we use the 2-block SBM to discuss this further. 
Each node has the probability $\pi_k$ of being in community $k$ for $k\in\{0,1\}$ and $\pi_0+\pi_1=1$. Let $\Omega_{uv}$ be the probability of an edge between nodes in group $u$ and $v$ for $0\leq u\leq v\leq 1$. Consider a adjacency matrix $A$ sampled from this model where $\bc^*$ are the true-labels and $\hat{\bc}$ are estimated labels, perhaps found using a procedure in Section \ref{sec:met}. We say that $\hat\bc$ {\it is strongly consistent} \cite{bickel2009nonparametric} for $\bc^*$ if
\begin{equation}\label{eq:cons}
    \lim_{n\to\infty}P(\hat\bc=\bc^*)\to 1
\end{equation}
where $n\to\infty$ means that the size (number of nodes) of the network is increasing.

Despite this being a foundational question, there is little work on it for the CP structure. Though not explicitly in the context of this structure, in \cite{snijders1997} general conditions are provided under which a procedure is consistent. One such approach is the following: let $d_{(1)},\dots,d_{(n)}$ be the (ordered) degrees of the nodes and let $I$ correspond to the index for which $d_{(I+1)}-d_{(I)}$ is maximal. Then assign $c_i=0$ if $i\leq I$ and $c_i=1$ if $i>I$. If $n_1/n\to\gamma_1$ where $n_1$ is the number of nodes in block 1 and $\gamma_0p_{01} + (1-\gamma_0)p_{00} < \gamma_0p_{11} + (1-\gamma_0)p_{01}$, then (\ref{eq:cons}) holds (see \cite{snijders1997} for details). In words, find the largest gap in consecutive, sorted degrees and then assign the nodes with smaller degrees to block 0 and those with larger degrees to block 1. This yields a simple procedure to consistently estimate the ground-truth labels. Further investigation is needed, however, to ensure that this result applies to the CP structure. Additionally, this a very specific procedure so it is unclear whether consistency holds for other algorithms. There is rich literature on consistency for community detection algorithms \cite{zhao2012consistency, amini2013, riolo2020, bickel2009nonparametric, mukherjee2021two}, so perhaps these ideas can be (carefully) extended to CP structures.
More work is needed to study the statistical consistency of the various CP metrics from Section \ref{sec:met}.


\subsection{Hypothesis testing}\label{sec:hyp}
Consider the following question: given a network, does it exhibit a statistically meaningful CP structure? This is the question of hypothesis testing for a CP structure. For any statistical hypothesis test, there are four key ingredients: (1) model parameter of interest, (2) a test statistic based on an estimator of this model parameter, (3) null model to establish `non-significance' and (4) rejection threshold to decide when to reject the null hypothesis. 

The biggest hurdles in networks hypothesis testing are determining the model parameter of interest (1) and deciding on a sensible null model (3). While (1) is not strictly necessary to carry out a test, it is crucial in order to root the test in the standard statistical hypothesis testing framework. None of the methods presented, save Zhang et al. \cite{Newmann2015}, mention the model parameter being estimated. This is not a merely a problem for the CP structure, but also in community detection hypothesis tests \citep[e.g.,][]{bickel2016:aa}. These methods simply present a test statistic without a model parameter. Additionally, a careful treatment of the null model is also lacking in the literature, although other authors have highlighted its importance \cite{Csermely2013}.  

\begin{figure}
    \centering
    \includegraphics[height=0.25\textheight, trim={2cm 2.5cm 2cm 2cm},clip]{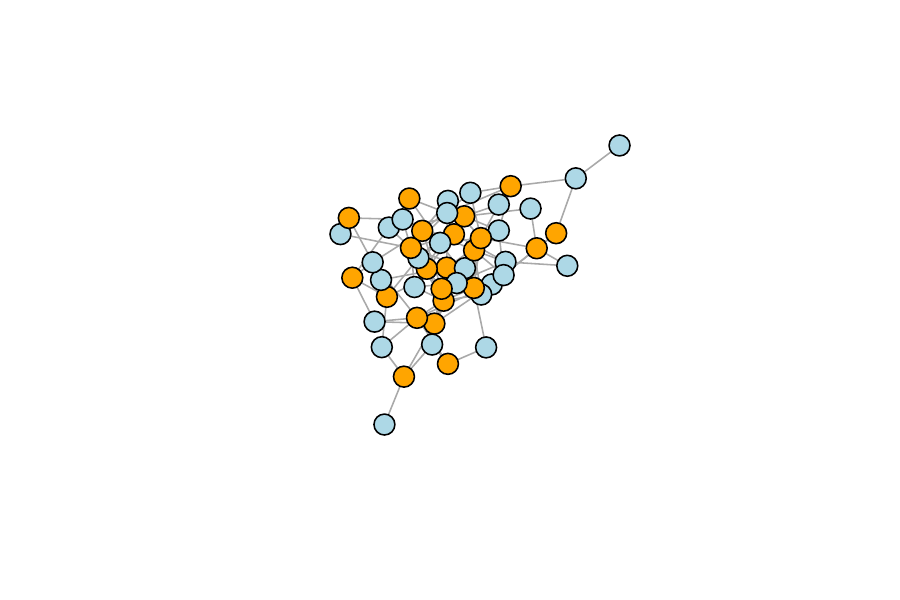}
    \caption{Erd\"{o}s-R\'{e}nyi network generated with $n=50$ and $p=0.10$. The method from \cite{BORGATTI2000} was used to find core-periphery labels where orange indicates a core node and blue indicates a periphery node. }
    \label{fig:motivate}
\end{figure}

Developing rigorous statistical methods to determine the presence of a CP structure is crucial to network science.
Otherwise we would likely claim that every network has a CP structure, which could be a false inference with significant practical implications. For example, consider a world trade network where nodes are countries and edges represent trade between the countries and assume that a recession occurs in one country.
If the network has a CP structure, then this recession would impact the entire world economy if it is a core node but only a minimal impact if it is a periphery node. 
As another example, consider a social network where nodes are people and edges are interactions, and assume that a highly contagious disease breaks out in a small set of people. If these people are part of a core, then the disease could easily spread to the entire network, but not if they are part of the periphery. As a further motivating example, consider the network in Figure \ref{fig:motivate}. This network was generated from an ER model with $n=50$ and $p=0.10$. The Borgatti and Everett \cite{BORGATTI2000} algorithm returned these CP labels where orange indicates a core node and blue indicates a periphery node. Since an ER model generated this network, these labels do not correspond to a meaningful CP structure. Thus, any investigation into similarities or characteristics of these ``core" nodes could be potentially misleading.

The first hypothesis test is from Boyd et al. \cite{BOYD2006}. The authors use $\rho$ in (\ref{eq:rho}) from Borgatti and Everett \cite{BORGATTI2000} to run a simple permutation test to yield a $p$-value. Specifically, the method computes the value of $\rho$ from the given network and then generates bootstrap samples from the network by randomly re-wiring edges in such a way that the total number of edges, $m$, is preserved but the degree distributions are not. Then the $p$-value is the proportion of bootstrapped networks which yield a larger value of $\rho$ than that of the original network. This routine is simple, easy to implement and relies on the most well-studied CP metric. 

The Boyd et al. test, however, fails to address key hypothesis testing ingredients. It does not specify the underlying model parameter being estimated. Additionally, the total number of edges is kept fixed during the bootstrap step, but this quantity is not an intrinsic feature of the network and thus may not be a reasonable feature of the network to use in the null. In other words, drawing another network from the same data generating process is unlikely to yield the same value of $m$. In \cite{kojaku2017} and \cite{elliott2020} the authors use a parametric bootstrap to generate ER networks with $p=\hat p=\sum_{i<j}A_{ij}/{n\choose 2}$ computed from the original network and then find the $p$-value empirically. This bootstrap step preserves the {\it expected} number of edges so the null model is similar to that of Boyd et al. The parametric bootstrap is a promising approach for hypothesis testing but the null model must be chosen with care. It is unlikely that any real-world network resembles an ER model so using it as the null model can lead to a high Type I error when a network is not ER but also does not have a CP structure {\it a la} the results shown in \cite{yanchenko2021}.

In Rossa et al. \cite{rossa2013} the authors present an informal test of significance. After computing the core-periphery centralization $C$ from (\ref{eq:rossa}), they generate 100 new networks with the same degree sequence as the original network. Then they compute $C^*_i$ using (\ref{eq:rossa}) for each randomized network $i=1,\dots,100$ and find the $z$-score,
\begin{equation}
    z
    =\frac{C-\text{mean}(C^*_i)}{\text{stdev}(C^*_i)}.
\end{equation}
A large $z$-score implies that the network has a significant CP structure. While this is a natural direction to follow, there is no theory establishing the validity of the normal approximation, and formal guarantees for the Type-I error or power of the test do not exist. Furthermore,  similar to Holme \cite{Holme2005}, this method uses the configuration model as the null. This approach, however, has known drawbacks. As many previous works highlight \cite{kojaku2017, Kojaku2018, elliott2020, kostoska2020core}, during the bootstrap step the high degree nodes are again assigned to the core which may lead to low power for the test. Taking this to the extreme, consider a network with an ideal CP structure $(A_{ij}=1$ if node $i$ or $j$ is in the core and 0 otherwise). The only network which preserves this degree sequence is the original network itself. Thus, there can be no re-wiring so any hypothesis test would return a $p$-value of 1, i.e., we would wrongly infer that there is no CP structure. In \cite{Kojaku2018} the authors propose adding a third block to the network to deal with this problem.

A fundamentally different approach to determining the significance of a CP structure (or more generally any mesoscale feature) comes from the {\it surprise} formulation \cite{aldecoa2013, de2019, marchese2021}. This paradigm assumes that there is a population of ${n\choose 2}$ node pairs of which $m$ (the number of observed edges) have been drawn. Of these $m$ realized edges, $M_c$ are ``successes" meaning that they are core-core or core-periphery edges. Then the $p$-value is the probability of observing as many or more ``successful" edges out of $m$ draws using the hyper-geometric distribution. These $p$-value computations are exact and do not require a bootstrap step.

\subsection{Bayesian Inference}\label{sec:bayes}
Bayesian inference on networks has seen great progress in areas such as estimation \cite{hoff2002latent, handcock2007model}, exponential random graph models \cite{caimo2011, thiemichen2016, tan2020} and community detection \cite{psorakis2011overlapping, morup2012bayesian, van2018bayesian}. Bayesian methods for the CP structure, however, are far less developed. We devote this subsection to the existing work in this area.

Snijders and Nowicki \cite{snijders1997} were perhaps the first to use Bayesian techniques with block models. This paper uses a two-block SBM likelihood simlar to (\ref{eq:like}) from Zhang et al. \cite{Newmann2015}. By requiring $p_{11}>p_{00}$, the parameters are identifiable. Then the authors propose the following Gibbs sampling procedure, starting at step $s$ of the sampler:
\begin{enumerate}
    \item Draw $(\gamma_1^{(s+1)},p_{11}^{(s+1)},p_{01}^{(s+1)},p_{00}^{(s+1)})$ from the posterior of $(\gamma_1,p_{11},p_{01},p_{00})|(A,\bc^{(s)})$
    \item Draw $c_1^{(s+1)}$ from $\bc_1|(\gamma_1^{(s+1)},p_{11}^{(s+1)},p_{01}^{(s+1)},p_{00}^{(s+1)}, A, c_2^{(s)},\dots,c_n^{(s)})$
    \item For $i=2,\dots,n-2$, draw $c_i^{(s+1)}$ from \\ $c_i|(\gamma_1^{(s+1)},p_{11}^{(s+1)},p_{01}^{(s+1)},p_{00}^{(s+1)}, A, c_2^{(s+1)},\dots,c_{i-1}^{(s+1)},c_{i+1}^{(s)},\dots,c_n^{(s)})$
    \item Draw $c_n^{(s+1)}$ from $c_n|(\gamma_1^{(s+1)},p_{11}^{(s+1)},p_{01}^{(s+1)},p_{00}^{(s+1)}, A, c_2^{(s+1)},\dots,c_{n-1}^{(s+1)})$
\end{enumerate}
Steps 2-4 leverage the fact that
\begin{multline}
    \frac{P(c_i=1|A, c_j, j\neq i)}{P(c_i=0|A, c_j, j\neq i)}\\
    =\frac{\gamma_1}{1-\gamma_1}(1-p_{11})^{-n_1}
    (1-p_{01})^{2n_1-n}(1-p_{00})^{n-n_1} \times \\ \left(\frac{p_{11}}{1-p_{11}}\right)^{-M_{1}}
    \left(\frac{p_{01}}{1-p_{01}}\right)^{M_{1}-M_{0}}
    \left(\frac{p_{00}}{1-p_{00}}\right)^{M_{0}}
\end{multline}
where $M_k=\sum_j A_{ij} \mathbb I(c_j=k)$ and $n_k$ is the number of nodes in group $k\in\{0,1\}$. The authors also show that if the block labels are known and vague priors are used for the edge probabilities, then the posterior distributions are:
\begin{multline}
    \gamma_1\sim\mathsf{Beta}(n_1+1,n_0+1),\ p_{11}\sim\mathsf{Beta}(M_{11}+1, \tfrac12n_1(n_1-1)-M_{11}+1)\\
    p_{01}\sim\mathsf{Beta}(M_{01}+1, n_1n_0-M_{01}+1),\ p_{00}\sim\mathsf{Beta}(M_{00}+1, \tfrac12n_0(n_0-2)-M_{00}+1).
\end{multline}
Although never expressly mentioned in their paper, this method clearly applies to estimating a CP structure. A major advantage of the Bayesian approach is that the posterior of $\bc$ gives the {\it probability} of each node being in the core as opposed to the {\it binary} assignment from a frequentist paradigm. If binary CP labels are required, then all nodes with a posterior mean of $c_i>0.5$ could be assigned to the core, for example.

Another paper that applies Bayesian methodology to study the CP structure is Gallagher et al. \cite{Gallagher2021}. The author's likelihood is the same as that of Snijders and Nowicki \cite{snijders1997}. This paper differs from previous work by focusing on the prior distribution for the block probabilities and explicitly enforcing the CP condition through this prior. In particular,
\begin{equation}
    P(p_{11},p_{01},p_{00})
    \propto \mathbb I(0\leq p_{00}\leq p_{01}\leq p_{11}\leq 1).
\end{equation}
The authors also propose a generalization of this prior to allowed for a ``layered" CP structure. Assuming there are $K$ blocks, or layers, the prior becomes
\begin{equation}
    P(p_{1},\dots,p_K)
    \propto \mathbb I(0\leq p_{K}\leq p_{K-1}\leq p_{1}\leq 1)
\end{equation}
where $p_{lm}=p_{\max\{l,m\}}$ is the probability of an edge between blocks $l$ and $m$. Lastly, a posterior odds ratio can be used to determine which type of CP structure is a better fit to the data. Mathematically, 
\begin{equation}
    \Lambda
    =\frac{P(\bc_b|A)}{P(\bc_\ell|A)}
\end{equation}
where $\bc_b$ are the labels for the block model and $\bc_\ell$ are the labels from the layered model. If $\Lambda>1$, then the block model gives a better fit and vice-versa if $\Lambda<1$.

The main contribution of Gallagher et al. \cite{Gallagher2021} is the data-driven comparison of two different formulations of the CP structure. This does not enforce a particular type of CP structure, rather, it allows the data to dictate the more likely structure. Conversely, the CP structure is assumed or ``hard-coded" into the model, meaning it does not yield a sense of the presence or absence of a CP structure, but only which type of CP structure is more probable. Still, this approach seems promising to apply to the hypothesis testing problem as well.

\section{Computation}\label{sec:comp}

Due to improved data collection mechanisms, the modern statistician now has access to massive amounts of data. For example, online social networks can have thousands \cite{rozemberczki2019}, millions \cite{backstrom2006} or even billions \cite{yang2015} of nodes and edges. Since the total possible number of core-periphery assignments increases exponentially with $n$, the number of nodes, if CP methods are to be relevant in our data science age, they must be computationally scalable. This is an important problem because even the authors of the well known UCINET package \cite{borgatti2002} (which includes the Borgatti and Everett \cite{BORGATTI2000} method) state that the computational routines are only fast enough for networks with 5000 nodes or less. This section discusses different computational techniques, including those for big data.

\subsection{Objective function maximization}
Most CP methods require maximizing an objective function over the set of candidate assignments whose size increases exponentially with $n$.
Since an exhaustive search for the maximum is computationally infeasible, heuristics and approximations must be employed.

A well-known strategy for objective function maximization is a {\it greedy algorithm} \cite[e.g.,][]{Kernighan1970}, where the locally optimal choice is selected at each iteration. This algorithm switches node labels between core and periphery and keeps the label which results in a larger value of the objective function. Many existing methods use variants of this approach \cite{BOYD2006, kojaku2017, Kojaku2018, shen2021}. This paradigm is simple and fast but can return local optima. One way to mitigate this risk is to consider several different starting values.

Another optimization paradigm is {\it genetic algorithms} \cite{anderson1994, sastry2005}. Taking their cue from micro-evolutionary concepts, these algorithms consider a ``population" of possible solutions where the ``fittest" solutions ``mate" to form new solutions. Here, a solution is a CP membership vector $\bc$, the fitness is the value of the objective function for that label, and mating is the merging of two labels. Borgatti and Everett \cite{BORGATTI2000, borgatti2002} implement this algorithm in their work. These methods are better than greedy algorithms at reaching the global maxima, but suffer from increased computational complexity. A large number of tuning parameters and slow speeds can also plague these algorithms. Other popular combinatorial optimization routines include simulated annealing \cite{kirkpatrick1983, van1987, dowsland2012} and tabu search \cite{glover1986future, glover1998tabu}. 
To the extent of our knowledge, there has not been a systematic comparison of these different algorithms on different CP quantifiers.

The previous procedures can, in theory, search the entire space of solutions. A fundamentally different approach which greatly restricts the search space, but increases speed, is what we coin as {\it node-ordering methods}. Here, some variable like degree or centrality is used to order the nodes. Then, starting with the first (largest value) node in the core, nodes are added to the core one-by-one in decreasing order and the objective function is computed at each iteration. After all nodes have been added to the core, the labels which correspond to the largest objective function are kept. Several authors propose this algorithm with ordering based on degree \citep[Lip,][]{lip2011}, centrality \citep[da Silva et al.,][]{da2008} or Core-Score \citep[Cucuringu et al.,][]{cucuringu2016detection}. This approach is significantly faster because it only considers $O(n)$ possible solutions as opposed to $O(2^n)$, but this smaller search spaces also means that the global maximum may not be achieved.

To see why this idea is appealing for large networks, consider Lip \cite{lip2011} which leverages this framework using the objective function (\ref{eq:brusco}) proposed by Brusco \cite{BRUSCO2011}. The author shows that, for a fixed core size $K$, (\ref{eq:brusco}) can be re-expressed as 
\begin{equation}\label{eq:lip}
    Z(A,\bc)
    =\left(\frac12\sum_{i}d_i+\frac{K(K-1)}{2}\right)
    -\sum_{i,c_i=1}d_i.
\end{equation}
The first term does not depend on the core so it follows that the core should be the $K$ nodes with largest degree. This generalizes to the case when $K$ is unknown by adding nodes to the core in decreasing order of degree and keeping the core which corresponds to the largest value of the objective function. This algorithm runs in $O(n^2)$ time, 
and finds the core of a network with $n=50,000$ in under one second. While the particular form of the objective function in (\ref{eq:lip}) lends itself to a node-ordering algorithm, in general, it is unknown when the node-ordering paradigm adequately approximates the global maximum.

Lastly, likelihood maximization allows for an even broader set of computational procedures. Since the CP labels can be thought of as missing or hidden data, the Expectation-Maximization (EM) algorithm \cite{dempster1977maximum} is a natural choice for likelihood maximization \cite[e.g.,][]{snijders1997,Newmann2015}. Bayesian methods like Markov Chain Monte Carlo \cite[e.g.,][]{geyer1991} or Variational Bayes \cite[e.g.,][]{kingma2013} are other popular choices when working with likelihoods. See Section \ref{sec:bayes} for a further discussion on Bayesian methods. As a case study, consider Shen et al. \cite{shen2021} which uses the likelihood-based approach to allow for multiple cores and multiple edges between a pair of nodes. Here we present a simplified version for a single core and binary edges. The authors consider a DCSBM to model the CP structure. If the probability of a core-core or core-periphery edge is $p_c$ and periphery-periphery edge is $p_p$, then the likelihood is
\begin{equation}\label{eq:shen}
    P(A|\bc,p_c,p_p)
    =\prod_{i<j, c_i=1\text{ or }c_j=1} (\theta_i\theta_j p_c)^{A_{ij}} e^{-\theta_i\theta_j p_c}
    \prod_{i<j, c_i=c_j=0} (\theta_i\theta_j p_p)^{A_{ij}} e^{-\theta_i\theta_j p_p}
\end{equation}
Then the authors propose a greedy algorithm to maximize (\ref{eq:shen}) and apply it to networks with up to $80,000$ nodes.

\subsection{Transport}
Transport-based methods admit an entirely new suite of computational tools. For example, Holme \cite{Holme2005} uses $k$-cores. We briefly describe a routine based on \cite{newman2018networks}. First, any node with degree less than $k$ cannot possibly be in a $k$-core so these nodes (and corresponding edges) are removed from the network. But this means that some of the remaining nodes will now have a smaller degree, so now again we remove any nodes with degree less than $k$. This process continues until all remaining nodes have degree $k$ and what remains is, by definition, a $k$-core. Even though this core definition is slightly restrictive, calculating a sequence of $k$-cores is computationally cheap ($O(m)$), which makes it appealing for large networks. Indeed, Holme applies this approach to a network with hundreds of thousands of nodes and over a million edges in \cite{Holme2005}. Other transport methods rely on measures like distance and betweenness centrality, which take $O(m+n)$ and $O(n(m+n))$ time, respectively. We omit discussions of these algorithms but refer the interested reader to \cite{newman2018networks}. Several other existing authors specifically address computational feasibility in large networks. Most approaches, however, are designed specifically for one particular CP descriptor. Therefore, an open problem is to develop techniques which work for an arbitrary metric as there is in the community detection literature, e.g. divide-and-conquer \cite{pujol2009, zeng2015, mukherjee2021two}.

\section{Real-world applications}
\label{sec:data}
Core-periphery structures have been observed in many empirical studies across disciplines. One of the most common applications where CP structures arise is in infrastructure networks. For example,\footnote{This example derives from a transport-based view of CP structures as it is based on shortest paths between nodes.} every traveler from a small city in the United States knows that airport networks have a CP structure since, to travel to most places, the first flight is to a large airport and then the second flight is the destination of interest. Alternatively, those who live near large airports can easily fly to other large airports as well as to small airports. In this case, large or ``hub" airports make up the core and all other airports comprise the periphery. This structure has been formally studied in \cite{kojaku2017}, \cite{lordan2017analyzing}, \cite{lordan2019core} and \cite{naik2021}. One of the interesting findings from \cite{naik2021} is that hubs for Southwest Airlines (MDW and DAL) do not appear in the core of their analysis which means that Southwest may not use the same route structures as other airlines, i.e., funnelling all flights to their respective hubs. 

Another field where CP structures are well-studied is economics. 
Krugman \cite{krugman1996self} famously studied how different economies arrange themselves in a CP structure and \cite{magone2016core} found similar economic structure specifically in the European Union (EU). These authors note how this naturally leads to unequal power dynamics, making it difficult for periphery countries to affect policy and financial decisions. Not surprisingly, then, the countries hardest hit by the Eurozone crisis were those in the periphery, e.g., Greece, Spain, Portugal, Italy. In \cite{tickner2013core}, the authors highlight a well-known fact in international relations that the vast majority of core countries are in the northern hemisphere and periphery countries are in the southern hemisphere, a different kind of imbalance associated with a CP structure. Lastly, the authors in \cite{fricke2015core} explore the CP structure of overnight money lending markets. With Italian banks as nodes and interbank positions 
as edges, the authors found that not only does this network exhibit a CP structure, but that the core shows little change over time. This phenomenon has also been observed in German banks \citep{craig2014interbank}. 

Many social networks also exhibit a CP structure.
In \cite{cattani2008}, the authors explore creativity in the film industry, with nodes representing crew members from the seven major Hollywood film studios (Universal, Paramount, Warner Bros, Columbia-Tristar, Disney, 20th Century Fox and MGM) and edges representing collaboration on the same movie.
They conclude that people in-between the core and periphery have the greatest creativity since they are close enough to the core to have sufficient influence but far away enough that they still maintain their unique, creative skills.  In Yang et. al \cite{yang2018structural}, the authors explore the relationship between the community and CP structure in Twitter data. For this network, a node is a Twitter user in Dubai and there is an edge between users if there is a ``retweet" or ``mention" of another user over a thirty day period. One of the main findings in this work is that, in this network with many communities, each individual community exhibits a CP structure. 

Other applications that have observed CP structures include: academic citations \cite{mullins1977group, doreian1985structural, zelnio2012identifying,sedita2020invisible, wedell2022center}, open source software development \cite{amrit2010exploring}, online discussion groups \cite{beck2003individual}, grieving widow's family structure \cite{morgan1997stability}, and political blogs \cite{adamic::2005aa, Newmann2015}. In the political blogs example (Zhang et. al \cite{Newmann2015}), this structure is unsurprising as all blogs tend to cite famous blogs (core) whereas relatively unknown blogs (periphery) are less likely to be referenced. This structure, however, may make it more difficult for a small blog to gain prominence as the majority of the focus is on established blogs.

While many statistics papers have included real-world applications, most, if not all, of the applied papers studying CP structures have come from other disciplines and are published in non-statistics journals. Thus, an open avenue for research is for statisticians to investigate interesting, complex, real-world networks for CP structures. For the interested reader, we provide the following public network data repositories as a starting point: \cite{DuBois:2008, snapnets, nr, networkdata, igraphdata}.

\section{Open Problems}\label{sec:open}
We conclude with some important open problems in statistical inference for CP structures.

\begin{enumerate}
    \item {\it How can the current CP structure methods be unified, specifically regarding a model parameter intrinsic to this structure?}
    
    Each method presented in Section \ref{sec:met} can be interpreted as an estimator that implicitly defines the CP structure in a unique way. 
    The majority of the existing techniques make no mention of an underlying model parameter. 
    Not only does this lack statistical rigor but it also makes it difficult to compare different methods. A major breakthrough would be identifying and estimating a model parameter that describes the CP structure in a network. Ideally, this parameter would not be tied to a specific model structure (like SBM) to allow for fair comparison across different models. Additionally, it should be interpretable in order to yield a succinct summary of the strength of the CP structure. \\

    \item {\it What is an appropriate null model to represent ``no CP structure"?}
    
    Without a satisfactory answer to this question, it is impossible to delineate between a ``true" CP structure and simply noise or some other feature in the network. To this end, many authors have suggested different null models, e.g., Erd\"{o}s-R\'{e}nyi (ER), re-wiring preserving total number of edges and re-wiring preserving degrees (configuration model). There has, however, been little justification for the choice as well as no systematic comparison of the different nulls. Furthermore, these simplistic null models can lead to ubiquity, meaning that CP structures seemingly appear in {\it all} real-world networks and thus making it a useless measure in practice.\footnote{For a discussion of this phenomena in other meso-scale features, see \cite{bassett2017small,dong2018resilience}.} We argue, however, that the CP structure is {\it not} ubiquitous but is mistakenly attributed to many networks because of simplistic benchmarks. For example, the ER model is a popular CP benchmark but it is almost tautological that real-world networks deviate from this model. It is this deviation that leads to the ``universality" of CP structures in real-world networks. Thus, having a nuanced and rigorously justified null model is paramount to advancing  hypothesis testing for CP structures.\\

    \item {\it What is the connection and/or trade-off between the CP and other meso-scale structures like communities?}
    
    It is unlikely that any real-world network {\it exclusively} exhibits a CP or community structure. Rather, it is more likely that networks are a mixture of different features, similar to the DCSBM which incorporates degree heterogeneity and block community structure. 
    The PABM proposed by \cite{senguptapabm} generalizes the DCSBM and allows for flexible core-periphery structure within communities.
    In this direction, \cite{Yang2014} argues that cores arise from the intersection of many overlapping communities. In \cite{tuncc2015unifying} the authors provide a unified formulation that allows for a hybrid of a community and CP structure. In particular, the edge probability, $p_{ij}$ of a node in group $i$ and $j$ is modeled as
    \begin{equation}
        p_{ij}
        =a\delta_{c_i,c_j}(C_i+C_j-C_iC_j) + b
    \end{equation}
    where $\delta_{xy}$ is the Kronecker delta taking value $1$ if $x=y$ and 0 otherwise, $c_i$ is the group of node $i$ and $C_i$ is some measure of ``coreness" for node $i$. Thus, the $\delta_{c_i,c_j}$ accounts for the community structure and $C_i+C_j-C_iC_j$ accounts for the CP structure. In \cite{yang2018structural} the authors found that Twitter networks are composed of multiple communities with a CP structure within each community. Lastly, in \cite{Kojaku2018}, the authors argue that a third block (like a community) is needed for a CP structure.\\
    
    \item {\it Multiple core-periphery structures}
    
    Related to the previous point, another important research direction is identifying multiple CP structures in a network. Many of the existing approaches (e.g., Borgatti and Everett \cite{BORGATTI2000}) are defined to identify a single core in the network. Real-world networks, however, have been shown to exhibit multiple cores and peripheries, as in Zhang et. al \cite{Newmann2015} and Yang et. al \cite{yang2018structural}, where CP structures are observed within communities. There has been some methodological work on this problem by Kojaku and Masuda \cite{kojaku2017} and Shen et al. \cite{shen2021}. Kojaku and Masuda extend the Borgatti and Everett \cite{BORGATTI2000} metric in (\ref{eq:rho}) to allow for multiple cores. The proposed algorithm also automatically estimates the number of CP structures in the network and yields the statistical significance of this structure by considering an ER null model.\\
    
    \item{\it Core-periphery identification in large networks}
    
    While a plethora of methods exist for community detection in large networks \cite[e.g.,][]{Blondel:2008aa, de2011, Ruan2013}, much less work exists for CP identification. In fact, the largest networks analyzed for a CP structure appear in Holme \cite{Holme2005} which used a restrictive definition. The method in Borgatti and Everett \cite{BORGATTI2000}, for example, is typically limited to networks with no greater than $n=5000$ nodes. Creating new algorithms, extending existing algorithms to larger networks, borrowing ideas from other network literature \cite[e.g., divide and conquer,][]{song2014influence, mukherjee2021two} or some combination of these is necessary for CP identification to be scalable to huge networks.\\
    
    \item {\it More complicated networks} 
    
    Most of the work done so far has been on relatively simple network structures. It would be interesting to investigate more complicated scenarios like temporal \citep{holme2015modern, holme2012temporal}, multilayer \citep{kivela2014multilayer, boccaletti2014structure} or heterogenous \citep{shi2016survey, wang2019heterogeneous}  networks as well as hypergraphs \citep{ghoshal2009random, ouvrard2020hypergraphs}.
    
    
\end{enumerate}

\begin{acks}[Acknowledgments]
We would like to thank the Editor, Associate Editor, and two anonymous referees for very helpful comments that significantly improved the manuscript's quality.
\end{acks}


\bibliographystyle{imsart-number} 
\bibliography{refs, ref, refs2}      


\end{document}